# RELIABILITY AND ADMISSIBILITY OF AI-GENERATED FORENSIC EVIDENCE IN CRIMINAL TRIALS

Sahibpreet Singh[*]
Lalita Devi[**]

## ABSTRACT

*This paper examines the admissibility of AI-generated forensic evidence in criminal trials. The growing adoption of AI presents promising results for investigative efficiency. Despite advancements, significant research gaps persist in practically understanding the legal limits of AI evidence in judicial processes. Existing literature lacks focused assessment of the evidentiary value of AI outputs. The objective of this study is to evaluate whether AI-generated evidence satisfies established legal standards of reliability. The methodology involves a comparative doctrinal legal analysis of evidentiary standards across common law jurisdictions. Preliminary results indicate that AI forensic tools can enhance scale of evidence analysis. However, challenges arise from reproducibility deficits. Courts exhibit variability in acceptance of AI evidence due to limited technical literacy and lack of standardized validation protocols. Liability implications reveal that developers and investigators may bear accountability for flawed outputs. This raises critical concerns related to wrongful conviction. The paper emphasizes the necessity of independent validation and, development of AI-specific admissibility criteria. Findings inform policy development for the responsible AI integration within criminal justice systems. The research advances the objectives of Sustainable Development Goal 16 by reinforcing equitable access to justice. Preliminary results contribute for a foundation for future empirical research in AI deployed criminal forensics.*

---

[*] PhD Research Scholar, School of Law, Lovely Professional University, Phagwara, Punjab.
[**] Assistant Professor, School of Law and Legal Studies, DAV University, Jalandhar, Punjab.

## I. INTRODUCTION

Artificial intelligence has rapidly permeated nearly every facet of modern criminal justice systems. It is particularly influencing forensic science. AI is transforming evidence collection and analysis in a variety of fields.[1] In digital forensics, algorithms retrieve and analyse data from electronic devices. AI is used in facial recognition systems. It aids in voice and handwriting pattern analysis. Predictive analytics are applied to reconstruct crime scenes.[2] This is possible due to AI's capacity to process enormous amounts of data and find hidden correlations. Police departments, prosecutors, and forensic specialists are using these technologies more and more to produce leads and even evidence in court.[3] The main legal and scientific dilemma as law enforcement implements these technologies is whether AI-generated forensic evidence can satisfy the accepted criteria of dependability and admissibility in criminal cases. Traditionally, courts have relied on human experts whose credentials, methods, and findings are subject to cross-examination. In contrast, jurors, judges, and attorneys find it challenging to understand how an algorithm arrived at a certain result because AI systems frequently function as "black boxes."[4] This paper explores the practical, ethical, and liability implications of relying on machine-generated outputs in criminal proceedings. The analysis is limited on admissibility and reliability rather than more comprehensive discussions of AI's ethics or governance at the policy level. Adoption of AI without sufficient validation and legal control may jeopardise due process rights and erode public trust in justice, despite its vast potential to improve forensic efficiency and precision.[5] UNESCO's 2021 Recommendation establishes that explainability and accountability are essential for AI legitimacy.[6] Therefore, this study is objected to find a path which strikes a compromise between technical innovation and long-term legal protections.

## II. AI IN CRIMINAL FORENSICS

### i. Overview of AI tools used in Criminal Investigations

Many AI-based tools are used in modern forensic practice to speed up and enhance investigation procedures. These algorithms use face recognition to identify criminals or missing people by comparing their recorded or real-time photos to an extensive database.[7] The police departments in the United States and the United Kingdom have employed systems like Clearview AI, which illustrates the potential and controversy of the various automated identification techniques.[8] AI algorithms help spot abnormalities and reconstruct human behaviour in digital forensics by helping investigators sort through terabytes of data, including emails, chat logs, social network posts, photos, and sensor data.[9] Digital platforms use machine-learning algorithms that can easily identify the communication patterns, detect picture alteration, and cluster linked files.[10] AI systems are used in predictive analytics and crime-scene reconstruction to simulate the events based on sensor inputs, ballistic trajectories, or digital traces. These systems often incorporate Bayesian networks or probabilistic reasoning to deduce expected outcomes.[11] Furthermore, deep-learning architectures are used for the analysis of spectral and spatial information in voice and handwriting identification. In order to identify suspects in terrorist or ransom crimes, AI-driven speaker recognition has been investigated.[12]

---

[1] Esra Nur Bal & Aylin Yalcin Saribey, *Artificial Intelligence in Forensic Science: Applications, Legal Framework and Criminal Liability Regime*, in 2025 INTERNATIONAL CONFERENCE ON ARTIFICIAL INTELLIGENCE, COMPUTER, DATA SCIENCES AND APPLICATIONS (ACDSA) 1 (2025).
[2] RECOMMENDATION ON THE ETHICS OF ARTIFICIAL INTELLIGENCE (2021).
[3] Paul Grimm, Maura Grossman & Gordon Cormack, *Artificial Intelligence as Evidence*, 19 NORTHWEST. J. TECHNOL. INTELLECT. PROP. 9 (2021).
[4] Ido Hefetz, *Integrating AI Systems in Criminal Justice: The Forensic Expert as a Corridor Between Algorithms and Courtroom Evidence*, 5 FORENSIC SCI. 53 (2025).
[5] Brandon L. Garrett & Cynthia Rudin, *Interpretable Algorithmic Forensics*, 120 PROC. NATL. ACAD. SCI. e2301842120 (2023).
[6] UNESCO, RECOMMENDATION ON THE ETHICS OF ARTIFICIAL INTELLIGENCE (2021).
[7] Jacqueline G. Cavazos et al., *Accuracy Comparison across Face Recognition Algorithms: Where Are We on Measuring Race Bias?*, 3 IEEE TRANS. BIOM. BEHAV. IDENTITY SCI. 101 (2021).

### ii. Benefits of AI Adoption

Using AI in forensic work has significant benefits. Machine-learning systems are notable for their speed. They can analyse large data sets in minutes. These can find connections that would take months for human analysts to find.[13] AI also promises consistency and objectivity. Once properly trained and validated, an algorithm applies uniform criteria to every case, potentially reducing human bias and fatigue.[14] In digital forensics, AI classifiers can tag similar image clusters uniformly, improving reproducibility.[15]

### iii. Risks and Limitations

Despite these advantages, bias, explainability, and data quality are major issues. Algorithmic prejudice is a persistent problem. According to a paper by Cuellar et al., when photos were degraded, top facial recognition algorithms misdiagnosed people from darker-skinned populations higher than those from lighter-skinned equivalents.[16] These prejudices may directly result in erroneous detentions or misdirected investigations. Results are further compromised by errors resulting from data-quality issues, such as incomplete or incorrectly labelled training sets.[17] The forecasts or correlations may not generalise when algorithms are programmed on datasets that are restricted by geography or demographic information. Buolamwini and Gebru's research revealed comparable differences in commercial face-analysis algorithms by gender and ethnicity.[18]

---

[8] Denise Almeida, Konstantin Shmarko & Elizabeth Lomas, *The Ethics of Facial Recognition Technologies, Surveillance, and Accountability in an Age of Artificial Intelligence: A Comparative Analysis of US, EU, and UK Regulatory Frameworks*, 2 AI ETHICS 377 (2022).
[9] Biodoumoye George Bokolo & Qingzhong Liu, *Artificial Intelligence in Social Media Forensics: A Comprehensive Survey and Analysis*, 13 ELECTRONICS 1671 (2024).
[10] Jeelani Ahmed & Muqeem Ahmed, *Classification, Detection and Sentiment Analysis Using Machine Learning over next Generation Communication Platforms*, 98 MICROPROCESS. MICROSYST. 104795 (2023).
[11] Abiodun A. Solanke, *Explainable Digital Forensics AI: Towards Mitigating Distrust in AI-Based Digital Forensics Analysis Using Interpretable Models*, 42 FORENSIC SCI. INT. DIGIT. INVESTIG. 301403 (2022).
[12] Bokolo and Liu, *supra* note 9.
[13] Stuart W. Hall, Amin Sakzad & Kim-Kwang Raymond Choo, *Explainable Artificial Intelligence for Digital Forensics*, 4 WIREs FORENSIC SCI. e1434 (2022).
[14] Roland Neil & Michael Zanger-Tishler, *Algorithmic Bias in Criminal Risk Assessment: The Consequences of Racial Differences in Arrest as a Measure of Crime*, 8 ANNU. REV. CRIMINOL. 97 (2025).
[15] Abiodun A. Solanke & Maria Angela Biasiotti, *Digital Forensics AI: Evaluating, Standardizing and Optimizing Digital Evidence Mining Techniques*, 36 KI - KÜNSTL. INTELL. 143 (2022).
[16] Maria Cuellar et al., Accuracy and Fairness of Facial Recognition Technology in Low-Quality Police Images: An Experiment With Synthetic Faces (May 20, 2025).

The explainability problem exacerbates these dangers. Even their designers frequently struggle to completely understand deep learning systems, especially convolutional neural networks.[19] This "black-box" aspect makes it highly challenging to examine in court, where opposing counsel must be able to question both the process and the outcome.[20] According to a 2022 National Institute of Standards and Technology (NIST) special publication, AI-based forensic conclusions may lose their scientific legitimacy if they are not interpretable.[21] Similar to this, the Artificial Intelligence Act 2024 of the European Commission classifies AI systems used for forensic and law enforcement applications as "high-risk". This classification requires human control. It also requires proper documentation. It further requires transparency.[22] However, real-time biometric identification systems in public spaces are classified as "unacceptable risk". They are largely prohibited. But have limited exceptions for law enforcement. Collectively, these various risks underscore that AI in forensics is not inherently reliable. Rather its probative value depends entirely on robust transparent validation. It further depends on continuous human intervention.[23]

### III. LEGAL RELIABILITY STANDARDS

Trials must be based on techniques that have been scientifically established and outcomes that are resistant to criticism.[24] The dependability of AI-generated outputs, such as image analysis, speech comparison, or predictive modelling, must be demonstrated to the same or even greater extent than that of conventional forensic methods when they are offered as evidence.[25]

#### a) The Daubert Standard

The standard for the admissibility of scientific evidence was set in the 1993.[26] This case was decided by the US Supreme Court. Trial judges act as gatekeepers under this standard. They evaluate the validity of the scientific process. They also assess the applicability of the process that supports expert testimony. The Daubert factors are as follows:

(1) testability;

(2) peer review and publication;

(3) known or potential error rates;

(4) the presence of standards governing the operation of the technology; and

(5) widespread acceptability within the relevant scientific community.[27]

---

[17] Neil and Zanger-Tishler, *supra* note 14.
[18] Joy Buolamwini & Timnit Gebru, *Gender Shades: Intersectional Accuracy Disparities in Commercial Gender Classification*, in PROCEEDINGS OF THE 1ST CONFERENCE ON FAIRNESS, ACCOUNTABILITY AND TRANSPARENCY 77 (2018).
[19] Itiel E. Dror, *Biased and Biasing: The Hidden Bias Cascade and Bias Snowball Effects*, 15 BEHAV. SCI. 490 (2025).
[20] Garrett and Rudin, *supra* note 5.
[21] REVA SCHWARTZ ET AL., TOWARDS A STANDARD FOR IDENTIFYING AND MANAGING BIAS IN ARTIFICIAL INTELLIGENCE (2022).
[22] Bal and Yalcin Saribey, *supra* note 1.
[23] Juan M. Durán et al., *From Understanding to Justifying: Computational Reliabilism for AI-Based Forensic Evidence Evaluation*, 9 FORENSIC SCI. INT. SYNERGY 100554 (2024).
[24] *Id.*
[25] Grimm, Grossman, and Cormack, *supra* note 3.
[26] *Daubert v. Merrell Dow Pharmaceuticals, Inc.,* 509 U.S. 579 (1993).

All of these standards presents complex challenges when it comes to AI-based forensics.[28] The quality and representativeness of the proprietary datasets used to "test" AI systems are frequently unknown to the court. Because commercial vendors protect algorithms as trade secrets, peer review may be restricted.[29] Model drift, training data, and environmental factors can all affect error rates. As a result, without thorough independent validation, many AI tools can fall short of the Daubert criterion.[30]

Innovative scientific methods must first prove their acceptability and dependability. The idea is consistent with AI evidence that the court cannot assess the algorithm's dependability in a meaningful way until the algorithm's design, dataset, and validation procedure are revealed.[31]

### b) The Frye Standard

The Frye test, which was derived from Frye v. United States[32], was the governing standard in many jurisdictions prior to Daubert. It stipulates that the evidence must come from a scientific principle or finding that is "sufficiently established to have gained general acceptance in the particular field." Although the Frye test is still valid in some states and other common-law nations, it poses special difficulties for AI evidence because there isn't a single professional agreement in the "particular field" of forensic AI.[33]

Artificial intelligence system is developing rapidly and frequently more quickly than professional associations are able to publish standards for accreditation. As a consequence, the popularity of a certain AI model does not necessarily translate into scientific acceptance. Therefore, in order to prevent popularity from being misinterpreted for dependability, courts must carefully balance innovation.[34]

---

[27] REFERENCE MANUAL ON SCIENTIFIC EVIDENCE: THIRD EDITION (2011).
[28] Gregory Schwartz, When Disciplines Disagree: The Admissibility of Computer-Generated Forensic Evidence in the Criminal Justice System (Dec. 24, 2024).
[29] Patrick Nutter, *Machine Learning Evidence: Admissibility and Weight*, 21 UNIV. PA. J. CONST. LAW 919 (2019).
[30] Surya Gangadhar Patchipala, *Tackling Data and Model Drift in AI: Strategies for Maintaining Accuracy during ML Model Inference*, 10 INT. J. SCI. RES. ARCH. 1198 (2023).
[31] Daniel Seng & Stephen Mason, *Artificial Intelligence and Evidence*, 33 SINGAP. ACAD. LAW J. 241 (2021).
[32] *Frye v. United States*, 293 F. 1013 (D.C. Cir. 1923)
[33] FRAIGMAN KAYE, SCIENCE LAW STANDARDS STATS: STANDARDS, STATISTICS AND RESEARCH ISSUES (2002).
[34] Stanley Greenstein, *Preserving the Rule of Law in the Era of Artificial Intelligence (AI)*, 30 ARTIF. INTELL. LAW 291 (2022).

### c) International Approaches to Reliability

In the United Kingdom, the Forensic Science Regulator's statutory Code of Practice (2023) requires that all forensic methods must be demonstrably valid through method validation. The technical records must be maintained to ensure reproducibility by competent practitioners.[35] Criminal Procedure Rules Part 19 requires expert evidence to be objective, unbiased, and sufficiently reliable for court admission.[36] The U.S. Department of Justice's December 2024 report[37] on AI in criminal justice emphasizes that AI-generated forensic evidence must meet established evidentiary standards for reliability, accuracy, and explainability. This particular to high-stakes applications. It is applicable in forensic analysis where errors can profoundly impact justice outcomes.[38] These advancements reflect a global legal agreement that an AI system's dependability cannot be assumed only because it functions well in controlled settings; instead, it must be empirically and procedurally shown in a courtroom setting.[39]

## IV. THE ADMISSIBILITY DILEMMA

The admissibility of AI-generated forensic evidence is closely linked to reliability but extends to procedural and constitutional dimensions. Apart from being reliable, evidence must also be pertinent, substantial, and acquired in accordance with due process. Courts must determine whether the admission of AI outputs that affect guilt determination violates the accused's right to a fair trial, which includes the rights to confront evidence and comprehend the foundation of conviction.[40]

### i. Authentication and Chain of Custody

The majority of evidentiary laws, such as the Federal Rules of Evidence (Rule 901) and the Indian Evidence Act (Sections 65A–65B), require digital or electronic records to be authenticated in order to demonstrate that they are legitimate and unaltered.[41] This entails confirming the data's provenance, the algorithm's integrity, and the analytical process's continuity in the case of AI evidence.[42] The U.S. Department of Justice has advised that forensic algorithms be treated as instruments whose calibration, software version, and audit logs must be preserved and documented.[43]

---

[35] Rune Kenneth Bauge et al., *Evaluating the Scope of Peer Review in Digital Forensics: Insights from Norway and the U.K.*, 65 SCI. JUSTICE 139 (2025).
[36] CODE OF PRACTICE (2023).
[37] ARTIFICIAL INTELLIGENCE AND CRIMINAL JUSTICE, FINAL REPORT, (2024).
[38] Grimm, Grossman, and Cormack, *supra* note 3.
[39] Sabine Gless, Fredric I Lederer & Thomas Weigend, *AI-Based Evidence in Criminal Trials?*, 59 TULSA LAW REV. 36 (2024).
[40] Rita Matulionyte et al., *Should AI-Enabled Medical Devices Be Explainable?*, 30 INT. J. LAW INF. TECHNOL. 151 (2022).
[41] Ashwini Vaidialingam, *Authenticating Electronic Evidence: Sec. 65B, Indian Evidence Act, 1872*, 8 NUJS LAW REV. 43 (2015).
[42] Janet Stacey et al., *A Responsible Artificial Intelligence Framework for Forensic Science*, 375 FORENSIC SCI. INT. 112548 (2025).
[43] FORENSIC TECHNOLOGY: ALGORITHMS STRENGTHEN FORENSIC ANALYSIS, BUT SEVERAL FACTORS CAN AFFECT OUTCOMES (2021).

AI-based evidence have been instances of dismissed by courts because of inadequate documentation. The Wisconsin Supreme Court affirmed the use of the COMPAS risk-assessment algorithm for sentencing in State v. Loomis, 881 N.W.2d 749 (Wis. 2016), although it also acknowledged transparency issues and cautioned that these systems cannot serve as the exclusive foundation for court decisions.[44]

### ii. Procedural and Constitutional Safeguards in Admitting AI Evidence

The admissibility of forensic evidence produced by AI presents foundational procedural issues. It also raises constitutional issues. These issues surpass basic scientific validity. Courts must protect the presumption of innocence. They must also protect due process. The defendant's right to a fair trial must be preserved. The right to a transparent trial must also be maintained. Automated decision-making technologies must not compromise these rights. The normative boundary that developing technologies must function within is formed by these protections.[45]

### iii. Transparency and Explainability

According to the right to be heard principle of *audi alteram partem*, the accused must be given a meaningful chance to contest the evidence used against them. This right is weakened when AI systems function as "black boxes," producing outcomes using opaque neural network topologies.[46] It becomes arduous to challenge an AI's dependability if neither the defence nor the court can comprehend the logic underneath its output.[47] The court specifically acknowledged this issue in State v. Loomis (Wisconsin Supreme Court, 2016), cautioning that the defendant's capacity to contest the scientific validity of the COMPAS algorithm was restricted due to its proprietary nature. The ruling established a warning precedent for striking a balance between efficiency and transparency. Even, if the evidence was permitted.[48]

### iv. Right to Confrontation and Cross-Examination

Every accused person has the right to confront witnesses. They also have the right to cross-examine witnesses. This right exists under the Sixth Amendment of the United States Constitution. It is also protected under Article 21 of the Indian Constitution. However, this right gets convoluted when AI systems function as algorithmic witnesses. A machine learning model cannot be cross-examined by defendants. Also, the developers may decline to reveal proprietary source code due to intellectual property concerns.[49] In an effort to lessen this conflict, courts have required explainability reports or algorithmic audits prior to admitting evidence based on artificial intelligence. For example in the case of Bridges v. South Wales Police [2020] EWCA Civ 1058 the Court of Appeal determined that the use of facial recognition technology violated privacy rules. It also violated equality rules. This was due to inadequate transparency. It was also due to a lack of safeguards. The Metropolitan Police of the United Kingdom was put under judicial review.[50] These cases highlight the need for AI-generated forensic evidence to meet constitutional fairness requirements. They also show the need to meet scientific reliability requirements for the evidence to be admitted.

---

[44] *State v. Loomis*, 881 N.W.2d 749 (Wis. 2016).
[45] Brandon L Garrett & Cynthia Rudin, *The Right to a Glass Box: Rethinking the Use of Artificial Intelligence in Criminal Justice*, 109 CORNELL LAW REV. 561 (2024).
[46] Chris Chambers Goodman, *AI, Can You Hear Me? Promoting Procedural Due Process in Government Use of Artificial Intelligence Technologies*, 28 RICHMOND J. LAW TECHNOL. 700 (2022).
[47] Bryce Goodman & Seth Flaxman, *European Union Regulations on Algorithmic Decision Making and a "Right to Explanation,"* 38 AI MAG. 50 (2017).
[48] *State v. Loomis*, 881 N.W.2d 749 (Wis. 2016).
[49] Danielle Citron & Frank Pasquale, *The Scored Society: Due Process for Automated Predictions*, 89 WASH. LAW REV. 1 (2014).

## V. LIABILITY AND ETHICS

### a) Attribution of Responsibility

When AI-generated evidence results in an erroneous conviction or acquittal of the accused persons, the dispersion of culpability among several players, including developers, law enforcement agencies, forensic specialists, and prosecutors, presents a unique challenge. Expert witnesses in traditional forensic science are accountable for the techniques they use.[51] In the United States, developers may be subject to product liability doctrine if an AI tool is determined to be unreasonably harmful in use or to have flawed design. However, if law enforcement organisations neglect to appropriately validate or interpret AI outputs, they may be considered responsible.

### b) Ethical Dimensions

Admissibility is further complicated by ethical issues. The impartiality of criminal adjudication can be compromised by algorithmic bias, especially racial or gender bias. Commercial facial recognition algorithms misclassified darker-skinned women at rates of up to 34.7%, compared to 0.8% for lighter-skinned men.[52] Article 14 of the Indian Constitution and the U.S. Fourteenth Amendment's protection of equality may be violated if biased algorithms generate evidence, such as voice matching or suspect identification. Evidence must be impartially produced in addition to being trustworthy in order to be considered due process. The idea of technical due process put out by academics is based on this moral precept.[53]

## VI. CONCLUSION AND RECOMMENDATIONS

In this contemporary era, artificial intelligence is rapidly transforming the criminal forensics technologies by offering significant unprecedented capabilities in data analysis, various pattern recognition, facial or voice identification, and crime-scene reconstruction. These advancements accelerate investigations while enhancing consistency. It potentially uncovers critical evidence that would otherwise remain hidden. However, there are many challenges with AI-generated evidence.

---

[50] *Bridges v. South Wales Police*, [2020] EWCA (Civ) 1058 (Eng.).
[51] Francesco Contini, Elena Alina Ontanu & Marco Velicogna, *AI Accountability in Judicial Proceedings: An Actor–Network Approach*, 13 LAWS 71 (2024).
[52] Buolamwini and Gebru, *supra* note 18.
[53] Nithya Sambasivan et al., *Re-Imagining Algorithmic Fairness in India and Beyond*, in PROCEEDINGS OF THE 2021 ACM CONFERENCE ON FAIRNESS, ACCOUNTABILITY, AND TRANSPARENCY 315 (2021).

Explainability is a major concern. Dependability is another concern. Legal professionals may find it challenging to fully comprehend the outcomes. Judges may also find it difficult. Jury members may struggle to scrutinise the outcomes. Algorithms have the potential to incorporate biases. They can reflect shortcomings in data quality. They may also function as opaque black boxes. A careful balance between technological innovation and fundamental legal principles is required for the admissibility of AI-based forensic evidence. The liability considerations further complicate this landscape, as the responsibility for flawed AI outputs may be distributed across developers, forensic experts, and law enforcement agencies, which may result into raising both legal and ethical questions about accountability and impartiality of the flawed AI evidences. AI has the ability to improve the criminal justice system despite these obstacles if it is used carefully and with full accountability and transparency. The important things to be considered are the standardisation of evaluation frameworks, open procedures, ongoing human supervision, and interdisciplinary oversight combining legal, ethical, and technical knowledge, for integrating AI into forensic practice. Courts and investigative organisations can use AI to their advantage while upholding the integrity of these evidence. This will further lead to bolstering public trust in justice by including these protections. In the final analysis, it can be concluded that AI-generated forensic evidence is a promising field. But on the same part its full potential won't be realised until technological developments are accompanied by strict legal requirements, moral awareness, and careful application, guaranteeing that innovation advances justice rather than subverts it. To ensure effective integration of AI in criminal forensics, the following policies are proposed:

1. Establish clear liability attribution.
2. Product liability doctrines should apply to AI forensic vendors if systems are unreasonably dangerous or poorly designed.
3. Wrongful conviction remedies must address AI-generated errors with the same rigor as traditional forensic error.
4. Develop standardized AI validation protocols for forensic tools.